\documentclass{mn2e}
\usepackage{epsfig}
\usepackage{times}
\begin{document}

\title[SS Cyg] 
{Late-outburst radio flaring in SS Cyg and evidence for a powerful kinetic output channel in cataclysmic variables}
\author[Fender \& Bright]
       {Rob Fender$^1$\thanks{email: rob.fender@physics.ox.ac.uk}, Joe Bright$^1$, Kunal Mooley$^2$, James Miller-Jones$^3$\\
       $^1$Astrophysics, Department of Physics, University of Oxford, Keble Road, Oxford OX1 3RH, UK\\
       $^2$National Radio Astronomy Observatory, Socorro, NM 87801 USA and Caltech, 1200 E. California Blvd., MC 249-17, Pasadena, CA 91125 USA\\
       $^3$International Centre for Radio Astronomy Research - Curtin University, GPO Box U1987, Perth, WA 6845, Australia)}
\maketitle

\begin{abstract}
Accreting white dwarfs in binary systems known as cataclysmic variables (CVs) have in recent years been shown to produce radio flares during outbursts, qualitatively similar to those observed from neutron star and black hole X-ray binaries, but their ubiquity and energetic significance for the accretion flow has remained uncertain. We present new radio observations of the CV SS Cyg with Arcminute Microkelvin Imager Large Array, which show for the second time late-ouburst radio flaring, in April 2016. This flaring occurs during the optical flux decay phase, about ten days after the well-established early-time radio flaring. We infer that both the early- and late-outburst flares are a common feature of the radio outbursts of SS Cyg, albeit of variable amplitudes, and probably of all dwarf novae. We furthermore present new analysis of the physical conditions in the best-sampled late-outburst flare, from Feb 2016, which showed clear optical depth evolution. From this we can infer that the synchrotron-emitting plasma was expanding at about 1\% of the speed of light, and at peak had a magnetic field of order 1 Gauss and total energy content $\geq 10^{33}$ erg. While this result is independent of the geometry of the synchrotron-emitting region, the most likely origin is in a jet carrying away a significant amount of the available accretion power.
\end{abstract}


\section{Introduction}
SS Cyg was one of the first systems containing an non-magnetic accreting white dwarf which was shown to have an associated variable radio source (K\"ording et al. 2008). This variable radio component appeared to connect to the properties and phase of the outburst in a way which was qualitatively similar to patterns observed in black hole and neutron star accretors in X-ray binaries (Fender et al. 2004; M\~unoz-Darias et al. 2014). The radio outbursts from the source have been shown to repeat (Russell et al. 2016), suggesting a strong coupling between accretion and ejection, as expected by analogy with X-ray binaries. Most recently, Mooley et al. (2017) reported an intensive monitoring campaign at 14--17 GHz with the AMI-LA radio telescope in which a very strong and rapid radio flare was observed, peaking at $\sim 20$ mJy in just $\sim 15$ minutes towards the end of the outburst. Such a late-time radio flare had not been previously observed (Russell et al. 2016).
Since the discovery of radio flaring from SS Cyg, variable radio emission has been observed from a significant number of other CVs, suggesting that conclusions drawn by the study of this system should be broadly applicable to other CVs (K\"ording et al. 2011; Coppejans et al. 2015, 2016).

\begin{figure*}
\epsfig{file=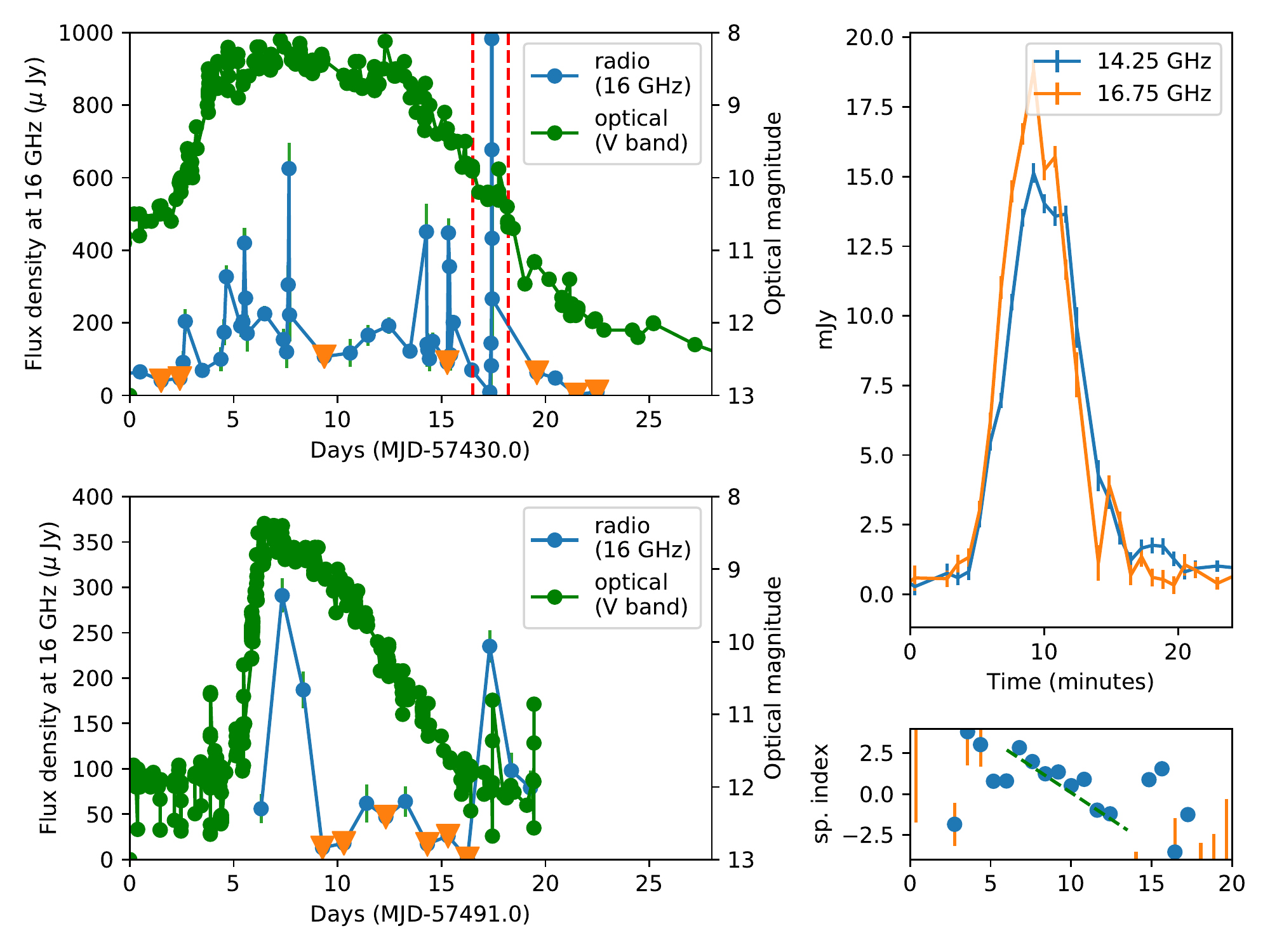, width=16cm}
\caption{{\bf Left panels:} Radio flux density from SS Cyg observed with AMI-LA in the 14--17 GHz band during the February (upper left panel) and April (lower left panel) 2016 outbursts. Optical magnitudes from AAVSO are overlaid. Note that the y-axis in Feb 2019 is cut at 1000 $\mu$Jy; the late-time flare was much brighter (see right panels). The light curves are qualitatively very similar, showing early and late-time radio flares separated by about ten days, with lower (but still sometimes detectable) levels of radio emission in between. The late-time flare in April 2016 is however much lower in peak flux density than that in February, suggesting that such ejection events are common but have a large range in amplitudes. Upper limits are indicated by the orange triangles. The red dotted lines in the Feb 2016 light curve highlight the flare which is isolated in the right panels. {\bf Right panels:} Optically thick radio flare from SS Cyg on 29 February 2016 (the late-time flare in the upper left panel). The upper panel shows the emission observed at 14.25 and 16.75 GHz by the AMI-LA radio telescope. The lower panel shows the spectral index between the two bands. The dashed green line in the lower panel highlights the strong evolution from optically thick (spectral index $\alpha > +0.5$) to optically thin ($\alpha \leq -0.5$) during the flare. This strong evolution, under the assumption it is due to varying self absorption to synchrotron emission, allows us to determine the magnetic field as function of source size at flare peak. This in turn allows us to estimate a minimum energy for the event of $L \geq 2 \times 10^{33}$ erg (see Fig 2). While spectral analysis of weaker flares is more difficult, all of the flare events, including the late-time flare in April 2016, are consistent with the same spectral evolution, and hence a powerful kinetic feedback channel.}
\label{ss}
\end{figure*}

\begin{figure}
\epsfig{file=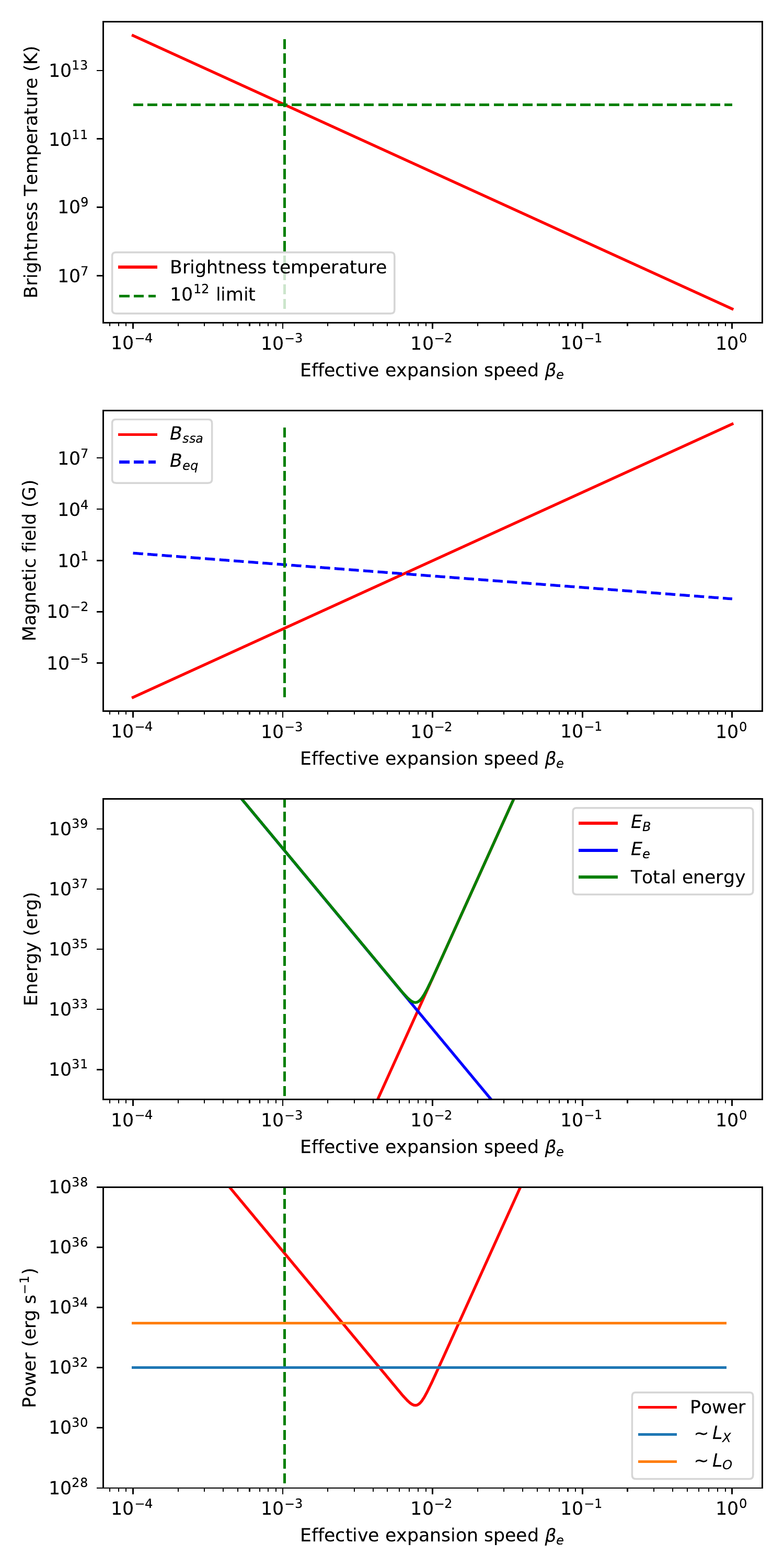, width=8cm}
\caption{Physical conditions in the synchrotron-emitting plasma associated with the optically thick radio flare. {\em Top panel:} Brightness temperature as a function of possible source size, indicating a minimum possible expansion speed of $\sim 10^{-3}c$ which would correspond to a maximum brightness temperature of $10^{12}$K. {\em Second panel}: magnetic field as a function of source size. {\em Third panel:} total minimum energy (particles plus field) in the ejecta as a function of source size, reaching a minimum at an expansion speed of $\sim 8 \times 10^{-3}c$ and corresponding minimum energy $\geq 2 \times 10^{33}$ erg. {\em Lower panel:} Power associated with the event, assuming energy input uniformly spread over rise time of the event, with a minimum power of $\geq 6 \times 10^{30}$ erg s$^{-1}$. Horizontal lines corresponding to the peak estimated X-ray ($L_X$) and optical ($L_O$) luminosity of the outburst are shown. }
\label{ss}
\end{figure}

For X-ray binaries containing black holes and neutron stars, the radio emission is commonly used as a way to measure the kinetic feedback from the accretion process (e.g. Fender \& M\~unoz-Darias 2016), which in some cases can be shown to exceed the power emitted directly in radiation (e.g. Fender, Gallo \& Jonker 2003). Jets from supermassive black holes in Active Galactic Nuclei can similarly carry away a large fraction of the available accretion power, which can have a fundamental impact on the formation and growth of galaxies (e.g. McNamara \& Nulsen 2012, Hardcastle et al. 2019).
Such measurements demonstrate the importance of understanding the jet in order to build a complete picture of the accretion process. However, such an approach has not yet been made for jets from white dwarfs in cataclysmic binary systems. This is due in part to the faintness of the radio emission compared to the X-ray binaries, but is mainly due to very few constraints on the size and hence total energy of the emitting region (see Fender \& Bright 2019 for a discussion of the very strong dependence of energy on physical size). 

Here we show, with new radio observations, that the late-time radio flaring of SS Cyg observed in Feb 2016 does not seem to be a unique occurence. We further re-analyse the large Feb 2016 flare, which showed clear radio spectral evolution consistent with a transition at the peak from optically thick to optically thin synchroton emission. We take the new approach to energy minimisation and characterisation of the physical characteristics of the ejecta, recently presented by Fender \& Bright (2019) and apply it to this event. 

\section{Observations}

\subsection{Radio} 

SS Cyg was observed extensively with the Arcminute Microkelving Imager Large Array (AMI-LA) radio telescope (Zwart et al. 2008, Hickish et al. 2017) in February and April 2016 in the 14--17 GHz band. The data and analysis of the February data are presented in Mooley et al. (2017); the reduction and analysis was the same for the April data. Most notable in Feb 2016 was a very strong radio flare event which occurred towards the end of the outburst and rose to over 20 mJy in less than 15 minutes, brighter than any previous event observed from this source (Russell et al 2016; although we note that for none of the outbursts has the observing cadence or spectral coverage been good enough that we can rule out having missed flare peaks). The observations in April 2016 also showed flares at the beginning and end of the outburst, qualitatively very similar to those in Feb 2016, although the amplitude of the late-time flare was much lower. These radio data are presented in Fig 1.

\subsection{Optical}

We use V-band optical data from the American Association of Variable Star
Observers (AAVSO) for our analysis. The data are plotted alongside the radio measurements in Fig 1. Interestingly, while the shape of the optical outburst in 2016 Feb was somewhat anomalous (slow rise), that in 2016 April was much more standard (see Cannizzo \& Mattei 1998).

\section{Analysis of the optically thick flare}

Under the assumption that the peak of the flare event is due to the evolution through optical depth unity to synchrotron self absorption, as implied by the spectral evolution (Fig 1, right panels), we are able to calculate the physical parameters of the synchrotron-emitting plasma as a function of size (expansion speed). Fig 2 presents the results of our analysis of the Feb 2016 late-time SS Cyg flare data, in which we use the approach of Fender \& Bright (2019) for a flare of amplitude 20 mJy at 16.75 GHz, a rise time of 300 sec and an assumed distance of 110 pc (Miller-Jones et al. 2013). Four panels are shown which compare the brightness temperature, magnetic field, energy and power as a function of the plasma expansion speed $\beta_e = v_{\rm exp} / c$.
Limiting the brightness temperature $T_B \leq 10^{12}$ K (from synchrotron cooling) sets a lower limit on the plasma expansion speeds of $1.0 \times 10^{-3}c$. The magnetic field derived from synchrotron self-absorption varies between $\sim 10^{-3}$ G at this minimum size to a completely unfeasible $\sim 10^{10}$ G if the expansion speed had been $c$. We also plot the magnetic field which would be derived assuming simple equipartition (the minimum energy if the size were known). This illustrates very clearly the issue with, say, assuming expansion of the ejecta at $\sim c$: the equipartition field for a source this size would be eight orders of magnitude too low to produce the observed synchrotron self-absorption. The third panel shows the minimum energy, which itself has a minimum which occurs for an expansion speed of $8 \times 10^{-3}c$ for which it has a value $2 \times 10^{33}$ erg and corresponding magnetic field of $3$ G (note the expansion speed is degenerate with the filling factor, but this does not affect the energy estimate - see Fender \& Bright 2019). Finally, the lower panel shows the variation of inferred power. The minimum power, assuming the kinetic energy was released over a period corresponding to the rise of the flare is $6 \times 10^{30}$ erg s$^{-1}$. 

\section{Discussion}

\subsection{Early- and late-outburst radio flaring}

Our radio observations of two outbursts from SS Cyg, in Feb and April 2016 respectively, suggest that both early- and late-outburst radio flaring are common features of the outbursts (despite the somewhat different optical outburst profiles). The amplitude of the early-outburst flare varies by at least a factor of two, while that of the late-outburst flare seems to vary by two orders of magnitude. It does seem likely that this reflects genuine physical variations in the synchrotron source, but we note again that -- particularly since the flaring can be so rapid and shows spectral evolution -- we may well have missed one or more peaks. Comparing our data with the VLA data compiled and discussed in Russell et al. (2016), it seems that the early-outburst radio flaring observed by both telescopes is likely to be the same. The late-outburst radio flaring observed twice with AMI, which occurs around 9 -- 11 days after the initial flare, could potentially have been missed by the VLA (see e.g. Russell et al. Fig 7 for comparison). While more data and better coverage are obviously desirable, we conclude that it is likely that both the early- and late-outburst radio flaring is a common, potentially ubiquitous, occurence in the outbursts of SS Cyg.
We will not discuss in any detail here how this flare may connect to the physical state of the accretion flow in SS Cyg, but we refer the reader discussions to in Wheatley et al. (2003) about late-outburst flaring and its association with transitions between UV- and X-ray dominated emission states. We do note in passing that for both outbursts reported here there does seem to be an optical brightening coincident with the late-outburst flare, but do not offer a physical interpretation.

\subsection{Kinetic power output channel}

We have established that the minimum energy associated with the optically thick radio flare from SS Cyg in 2016 Feb was $2 \times 10^{33}$ erg. We stress that this is very much a lower limit as it does not include any synchrotron luminosity beyond the observed bands, does not include any energy contribution from non-radiating particles (e.g. protons), and does not consider kinetic energy of bulk motion (although we do expect jets from CVs to have much lower bulk velocities than those in black hole and neutron star sources). Furthermore, if the rising phase, as expected, is due to expansion of the source, then at the earlier times the source would have had more internal energy. For example, in the early adiabatic expansion model of van der Laan (1966), the flux in the optically thick rising phase is proportional to source size as $F_{\rm thick} \propto r^3$.  The internal energy of such a source falls as $E \propto r^{-1}$, and so $E \propto F_{\rm thick}^{-1/3}$.
Looking at Fig 1, we can estimate that from the first detection of the flare rise to the peak, the source flux increased by about a factor of fifteen, which corresponds to a size increase by a factor $\sim 2.5$.
Hence the energy in the ejecta at the moment the flare was first detected was $\geq 5 \times 10^{33}$ erg.
The source may well have started considerably smaller, and with more internal energy, but it is impossible to tell from our data. The minimum energy size corresponds to ejecta with a radius of $\Delta t \beta_e c = 2.4$ light seconds, about half the binary separation for SS Cyg (Bitner, Robinson \& Behr 2007). If the ejecta started off initially on a size scale comparable to the white dwarf, that is around two orders of magnitude smaller than the size associated with the peak, the total energy content could have been $\geq 10^{35}$ erg, but this is very uncertain. Of course this naive approach does not take into account {\em in situ} particle acceleration by the conversion of kinetic energy to particles and field.

The instantaneous X-ray luminosity of SS Cyg at the time of the radio flare was probably around a few times $10^{31}$ erg s$^{-1}$ (based on observations of previous outbursts), and has been observed to reach up to $\sim 4 \times 10^{32}$ erg s$^{-1}$ (Wheatley, Mauche \& Mattei 2003, Mooley et al. 2017, Russell et al. 2017). The total radiative output (traced by the optical emission) is likely to peak at a few times $10^{33}$ erg s$^{-1}$ (Wheatley et al. 2003). 

Noting further that the time interval over which the $2 \times 10^{33}$ erg was injected via the kinetic energy channel into the synchrotron source is likely to have been {\em shorter} than the $\sim 300$s rise time of the event (which is instead dominated by the optical depth evolution of the source), the power into the synchrotron plasma was likely to be at least 10\% of the instantaneous X-ray luminosity, and at least 1\% of the total radiative output, possible considerably more. This is the first time a strong lower limit can be put on the power input into a synchrotron source from a CV. Note that the radio luminosity alone $L_R \sim 4 \pi d^ 2\nu F_{\nu} \sim 10^{27}$ erg s$^{-1}$, more than three orders of magnitude below the kinetic power constraint, and is not a good tracer of it. We stress that the analysis performed here does not assume a jet geometry or any bulk motion, so alternatives such as a spherical shell, are completely feasible and do not change the energy calculation. Regardless of geometry, it is unambiguous that the kinetic feedback responsible for the synchrotron emission is a very large fraction of the accretion power.

Besides the major flare event, most of the radio emission observed over the $\sim 20$ days of the outburst in Feb 2016 was in the form of short radio flares with amplitudes of a few 100 $\mu$Jy at 16 GHz and rise times of $\sim 10$ minutes. The spectral evolution of these events is not as well traced as for the stronger 20 mJy flare, but is consistent with evolution from optically thick to optically thin (see further discussion in Mooley et al. 2017). The late-time radio flare in April 2016 is also consistent with this spectral evolution. Using the same analysis as for the late-time Feb 2016 flare, the energy associated with these smaller flares is in the range $10^{31}$--$10^{32}$ erg, one to two orders of magnitude below that of the large flare. For these smaller flare events the expansion speed for minimum energy is similar to that calculated for the giant flare, although the magnetic field is smaller. This indicates a time-averaged kinetic power of $\geq 10^{28}$--$10^{29}$ erg s$^{-1}$ for $\geq 75$\% of the outburst (there is a $\sim 5$-day  phase in the middle of the outburst when the radio activity is more steady, and the methods applied here are not applicable).

Given the qualitative similarities in the coupling between accretion, in particular outbursts, and the occurence of radio emission in X-ray binaries and CVs it has been suggested that the radio emission from CVs originates in a jet (e.g. K\"ording et al. 2008, Russell et al. 2016 and references therein). In X-ray binaries, and AGN, the jet power has been estimated to be $\geq 10$\% of the instantaneous radiative luminosity during flaring states, and possibly to dominate over radiation as the main power output channel at lower accretion rates. Drawing these pieces of evidence together, our analysis strongly suggests that SS Cyg has a powerful jet during at least some phases of outburst, similar to the more relativistic systems. We note that VLBI observations of SS Cyg (Miller-Jones et al. 2013, Russell et al. 2017) report marginal evidence of a resolved jet of physical size $\sim 0.5$ A.U., from hints that the emission is resolved out on longer baselines. They note that the launch time for these ejecta should have been $\leq 2$ days earlier, since the previous VLBI images did not show a resolved component. A physical size of 0.5 A.U. two days after launch corresponds to a mean expansion speed of $\ga 10^{-3}c$, consistent within uncertainties to that we have derived from our optical depth analysis. It is interesting that our physical expansion speed estimate from the synchrotron self absorption condition is similar to those derived for X-ray binaries using both the same method of Fender \& Bright (2019), and also via direct fitting of a variant of the van der Laan (1966) model (Tetarenko et al. 2017).

Our physical analysis is the first robust estimate of the power in the kinetic feedback channel for a dwarf nova, or any cataclysmic variable. It indicates that during one large radio flare event the kinetic feedback channel was significant.  This  demonstrates quantitatively for the first time that kinetic feedback for dwarf novae is at times an important power output channel. Given our relatively poor sampling (compared to more radio-loud systems such as X-ray binaries), it is unlikely that this was a completely isolated event. Indeed observations of the next radio outburst reveal a similar, although lower-amplitude, late-time radio flare. Furthermore, $\sim$75\% of each outburst was characterised by smaller radio flares with a time-averaged power of at least 0.1\% of the instantaneous radiative luminosity, indicating a long phase of strong and powerful kinetic feedback, with a total kinetic feedback during the outburst of $>10^{35}$ erg, possibly much greater. Further high-cadence radio observations of SS Cyg and other dwarf novae, with sufficient bandwidth and sensitivity to measure spectral evolution, are crucial to understand how widespread this phenomenon is. 

\section*{Acknowledgements}
We thank the Mullard Radio Astronomy Observatory staff for scheduling and carrying out the AMI-LA observations. The AMI telescope is supported by the UK Science and Technology Facilities Council, the University of Cambridge and by  the European Research Council under grant ERC2012-StG-307215 LODESTONE. JB acknowledges funding from the UK Science and Technology Facilities Council. KM and RF acknowledge support from the Hintze foundation. 
JCAM-J is the recipient of an Australian Research Council Future Fellowship (FT140101082), funded by the Australian government. We acknowledge with thanks the variable star observations from the AAVSO International Database contributed by observers worldwide and used in this research. We thank Patrick Woudt for comments on a draft of this paper. We thank an anonymous referee for useful comments.

\appendix
\section{Minimum energy for a source of uncertain size}

It is well known that a minimum internal energy (and corresponding magnetic field) can be calculated for a synchrotron emitting plasma (Burbidge 1956; Pacholczyk 1970). The main problem in making the minimum energy estimate based on a spatially-unresolved radio flare event is the size of the source, which is extremely uncertain and yet has a very effect on the inferred minimum energy.

Fender \& Bright (2019) have recently shown, however, that a minium energy can be calculated, with associated size, if the peak of the light curve is due to synchrotron self absorption (similar but more extensive analysis for gamma-ray bursts has been presented by Barniol Duran, Nakar \& Piran 2013). Upper and lower limits on the possible range of sizes can be set by assuming that the expansion speed, $\beta_e$, is less than the speed of light but larger than the condition that the brightness temperature of the source exceeds $10^{12}$K. Under the assumption of approximately unity optical depth at flare peak, the magnetic field inside the plasma varies with the expansion speed as $B \propto \beta_e^{-4}$. As the magnetic field decreases, the energy in the electrons must increase. In fact a very strong minimum in the (minimum) internal energy as a function of size results from the $E_e \propto \beta^{-6}$ and $E_B \propto \beta_e^{11}$ dependencies of the total energy in electrons and magnetic field respectively. The equations governing the variation of the derived physical parameters as a function of $\beta_e$, as presented in Fig 2 of the main paper are:

\noindent
Brightness temperature

\[
T_B = \left[ \frac{F_{\nu} D^2}{2 \pi k_B \nu^2 \Delta t^2} \right] \beta_e^{-2}
\]

\noindent
Equipartition magnetic field

\[
B_{eq} = \left[ \left( \frac{9}{2} c_{12} L \right)^{2/7} (c \Delta t)^{-6/7} \right] \beta_e^{-6/7}
\]

\noindent
Magnetic field from synchrotron self absorption

\[
B_{ssa} = \left[ k_1 c^4 F_{\nu}^{-2} \left( \frac{\Delta t}{D} \right)^4 \nu_{\tau=1}^5 \right] \beta_e^4
\]

\noindent
where $F_{\nu}$ is peak flux density, $D$ is distance, $k_B$ is Boltzmann's constant, $\nu$ is observing frequency, $\Delta t$ is the observed rise time of the event, $\beta_e$ is the expansion speed of the plasma as a fraction of the speed of light, $L$ is the integrated radio luminosity, $c_{12}$ is a pseudo-constant from Pacholcyzk (1970) and given below, $c$ is the speed of light, $k_1 = 3.3 \times 10^{-61}$, $\nu_{\tau=1} = 0.7 \nu_{\rm peak}$ is the frequency at which the optical depth to synchrotron self absorption is unity.

The energies in the two components, particles and field, can then be calculated based upon the size and magnetic field in the ejecta, using:

\noindent
Energy in electrons

\[
E_e = \left[ c_{12} L \left( k_1 c^4 F_{\nu}^{-2} \left( \frac{\Delta t}{D} \right)^4 \nu_{\tau=1}^5 \right)^{-3/2} \right] \beta_e^{-6}
\]

\noindent
Energy in magnetic field

\[
E_B = \left[ \frac{c^3 \Delta t^3}{6} \left( k_1 c^4 F_{\nu}^{-2} \left( \frac{\Delta t}{D} \right)^4 \nu_{\tau=1}^5 \right)^{2} \right] \beta_e^{11}
\]

The sum of $E_e + E_B$, which is the minimum total energy, has a minimum as a function of $\beta_e$.

Finally the power is simply the energy divided by the rise time of the event. Since the rise time is likely to be much longer than the real power injection phase, this is again a lower limit. For complete details and a full discussion, see Fender \& Bright (2019).

The pseudo constant $c_{12}$ (Pacholcyzk 1970) depends upon the upper and lower frequency bounds and slope of the observed synchrotron emission:

\[
c_{12} = c_2^{-1} c_1^{1/2} \tilde{c}(p, \nu_1, \nu_2)
\]

where

\[
\tilde{c}(p, \nu_1, \nu_2) = \frac{(p-3)}{(p-2)} \frac{\nu_1^{(2-p)/2}-\nu_2^{(2-p)/2}}{\nu_1^{(3-p)/2}-\nu_2^{(3-p)/2}}
\]

and \smallskip
\noindent
$c_1 = \frac{3 e}{4 \pi m^3 c^5} = 6.27 \times 10^{18}$

\smallskip
\noindent
$c_2 = \frac{2 e^4}{3 m^4 c^7} = 2.37 \times 10^{-3}$

where $e$ is the charge on the electron, $m$ is the mass of the electron and $c$ is the speed of light.

Note that in the case the {\em filling factor} $f$ of the synchrotron-emitting plasma is less than unity, then this is absorbed in to (or can be considered to be degenerate with) the effective expansion speed as $\beta_e = (v_{\rm exp} f^{1/3})/{c}$.

\bsp

\end{document}